# Non-parallel voice conversion based on source-to-target direct mapping

*Sunghee Jung, Youngjoo Suh, Yeunju Choi, and Hoirin Kim*

School of Electrical Engineering, KAIST, Daejeon, South Korea

`{sh.ee, yjsuh, wkadldppdy, hoirkim}@kaist.ac.kr`

## Abstract

Recent works of utilizing phonetic posteriograms (PPGs) for non-parallel voice conversion have significantly increased the usability of voice conversion since the source and target DBs are no longer required for matching contents. In this approach, the PPGs are used as the linguistic bridge between source and target speaker features. However, this PPG-based non-parallel voice conversion has some limitation that it needs two cascading networks at conversion time, making it less suitable for real-time applications and vulnerable to source speaker intelligibility at conversion stage. To address this limitation, we propose a new non-parallel voice conversion technique that employs a single neural network for direct source-to-target voice parameter mapping. With this single network structure, the proposed approach can reduce both conversion time and number of network parameters, which can be especially important factors in embedded or real-time environments. Additionally, it improves the quality of voice conversion by skipping the phone recognizer at conversion stage. It can effectively prevent possible loss of phonetic information the PPG-based indirect method suffers. Experiments show that our approach reduces number of network parameters and conversion time by 41.9% and 44.5%, respectively, with improved voice similarity over the original PPG-based method.

**Index Terms**: voice conversion, non-parallel voice conversion, online inference, phonetic posteriogram

## 1. Introduction

The goal of voice conversion(VC) is to modify the para-/non-linguistic features of speech while retaining linguistic information to convert the voice of source speaker to that of target speaker. Voice conversion can be applied to any domain where speech communications take place. Those applications include AI speakers and robots, foreign language education, movie dubbing and speaking aids for speech-impaired patients, etc. There are two kinds of voice conversion approaches depending on the match of the source and target training corpora. One approach is the parallel method. One of the most earliest and still powerful study on this approach uses the Gaussian Mixture Model (GMM) as its mapping function between source and target speaker features [1]. The mapping is learned after aligning source and target features with dynamic time warping on the parallel corpus. The optimization of the mapping function is obtained by minimizing the mean squared error between target and converted features. The other approach is non-parallel voice conversion. Earlier methods of this approach used the iterative combination of a nearest neighbor search step and conversion step alignment (INCA) for obtaining alignment of non-parallel corpus [2]. This approach is composed of two steps. In search step, it aligns source and target features using the nearest neighbor algorithm. In conversion step, the mapping function from aligned source to target features is learned. It iterates these two steps until the transformation function converges. The learning steps were significantly time and memory consuming due to the nearest neighbor steps and its iterative alignment. In recent years, non-parallel voice conversion techniques use TTS or phone recognizer for obtaining alignment [3].

In the non-parallel approach utilizing TTS for alignment, the parallel corpus is generated first using the TTS system. Then the mapping from the source to target speakers is learned just as in the parallel approach [4, 5]. Another non-parallel approach employs phone recognizer for its alignment and is more widely used [6, 7]. This approach adopts the feature called phonetic posteriogram (PPG) which is a soft-labeled phone information used as a linguistic bridge between source and target acoustics. A lot of approaches stemmed from this method and they were shown to be effective on voice conversion tasks [7-9].

Regarding voice conversion architecture, parallel-voice conversion has some advantage over non-parallel approach. That is because the former approach employs only a single network for mapping the source to target acoustic features at conversion stage while the latter method based on PPG needs two networks. One of them is used for converting source features to PPGs and the other for converting PPGs to target features. The PPG-based non-parallel voice conversion approach has increased its usability since it does not need the parallel corpus. In this paper, we propose a new non-parallel voice conversion method which combines the strengths of both parallel and non-parallel approaches.

The proposed voice conversion method for improving conversion time/parameter efficiency is presented as follows.

In section 2, we describe the conventional non-parallel voice conversion approach using PPGs. In section 3, the proposed method is explained in detail. In section 4, experiments to prove the effectiveness of proposed methods are described. Finally, our conclusion is given in section 5.

## 2. Baseline non-parallel voice conversion

The baseline VC method introduced PPG as the speaker-independent bridging feature between source and target speakers in non-parallel VC tasks [6]. There has also been other approach that uses electromagnetic articulography (EMA) as its speaker-independent feature for this task [10]. However, taking into account the difficulty in obtaining EMA features, adopting PPGs for VC tasks provides a breakthrough for the usability of non-parallel VC.

The architecture of the baseline VC system is illustrated in Figure 1. As shown in this figure, the baseline approach has to train two networks in the training stages and run both of the networks in the conversion stage as well. In the training stage 1, the speaker-independent phone recognizer is trained on a multi-speaker corpus. Rather than adopting hard label from this phone recognizer, soft label called PPG, representing probability for all possible phones is adopted for more accurate conversion. In the second training stage, the nonlinear mapping function between PPGs and target speaker acoustic features is learned. In the conversion stage, which takes place on end-user devices, the source speaker utterance is transformed into PPGs by the phone recognizer aforementioned in train stage 1. Then the PPGs are transformed to target speaker features by the network trained in stage 2. Therefore, two networks have to be deployed on consumer devices, which makes it less efficient to convert in real-time. This problem can be alleviated by the proposed VC approach that employs a single network at run-time.

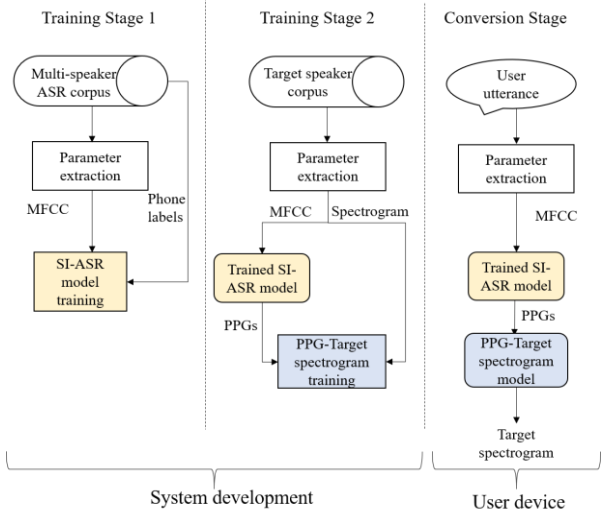

Figure 1: *Architecture of baseline non-parallel voice conversion.*

## 3. Proposed Methods

### 3.1. Overview of architecture

Figure 2 illustrates the proposed VC architecture for improving conversion step efficiency. Training stages 1 and 2 are the same as in the baseline architecture. Training stage 3 has been newly added. In training stage 3, we introduce a new network for source-to-target direct mapping. This new network is referred to as network 3 for the rest of this paper.

Conversion stage is the only stage that takes place in the user environment. Therefore, only this third network is deployed on the user device. This reduces memory requirement and conversion time.

### 3.2. Training stage 3

In training stage 3, mapping between multi-speaker MFCC and target linear spectrogram is learned. The network3 makes the conversion stage as simple as that in the parallel voice conversion approach. This network is trained with multi-speaker DB and it can be applied to any source speaker that is given in the conversion step. Therefore, the source speaker independence of the baseline PPG-based non-parallel voice conversion is kept in the proposed architecture as well.

Phone recognizer (network 1) from training stage 1 and PPG-to-target linear spectrogram network (network 2) from training stage 2 work together as an aligner between source and target speaker features. As a result, network 3 has aligned MFCC and linear spectrogram as input and output features. Aligned MFCC and linear spectrogram can have small covariate shift which is thought to be important factor in fast network learning [11, 12]. Therefore, the network 3 has negligible training cost compared with networks 1 and 2.

### 3.3. Implementation details

Network details are shown in Figure 3. Networks 1, 2, and 3 are implemented with the same structure. Inputs for each network are forwarded through a fully connected network, convolution bank, highway network and bidirectional gated recurrent network (CBHG) [13, 14] and a fully connected network at last for obtaining outputs of desired dimensionality. The only difference among the networks 1, 2, and 3 lies in the input and output features. Network 1 predicts PPGs from MFCCs. Network 2 predicts the probabilistic distribution of target speaker linear spectrograms given PPGs under the assumption of Gaussian mixture distribution for the output [15]. Network 3 takes MFCC as input and predicts the probabilistic distribution of target speaker linear spectrogram. Network 3 is trained with multiple speakers and can be applied for any source speaker in the conversion time. Here, no encoder-decoder model was adopted since encoder-decoder structure introduces additional network and increases both conversion time and number of parameters. The probability distributions of output target linear spectrograms in networks 2 and 3 are given in (1). These networks attempt to predict means, variances and mixture weights for linear spectrograms. The loss function is negative log likelihood as (2),

$$Pr(x_t|y_t) = \sum_{j=1}^{M} \pi_t^j N(x_t|\mu_t^j, \sigma_t^j) \quad (1)$$

$$L(x) = -\sum_{t=1}^{T} \log Pr(x_t|y_t) \quad (2)$$

where $y_t$ is PPGs for network 2 and normalized MFCCs for network 3, respectively. $x_t$ is linear spectrogram for target speaker. M is the number of mixtures. $\mu_t^j, \sigma_t^j$ and $\pi_t^j$ each represents mean, variance and mixture weight for j th mixture at time t [15].

## 4. Experiments and results

Network 1 was trained on the TIMIT corpus for once and was shared for all gender-pair experiments [16]. The VCTK corpus was used for training networks 2 and 3 [17]. Two females and two males were randomly chosen without any constraints on their accents. Network 2 was trained on one female and one male target. Network 3 was trained on the dataset composed of 105 speakers excluding the 4 speakers who are used either as source or target. The total number of networks to be trained for four-pair voice conversion experiments was five. That is, one for network 1, two for network 2, two for network 3 (from multi-speaker corpus to female and male target each). A three-layered fully connected network with dropout rate of 0.2 preceded CBHG. CBHG was constructed with 512 hidden units, 8 filter banks, 8 highway networks and bi-directional GRU with 512 units. A single-layered fully connected network followed CBHG to transform the output of CBHG into desirable dimension for each network 1, 2 and 3. Speech data are sampled at 16 kHz. Input features are composed of 401 frames. The outputs of network 2 and 3 were modeled with GMM of 5 mixtures to predict 257-dimension linear spectrogram.

In the performance evaluation, firstly, Mel-cepstral distortion (MCD) was measured upon all possible gender-pairs

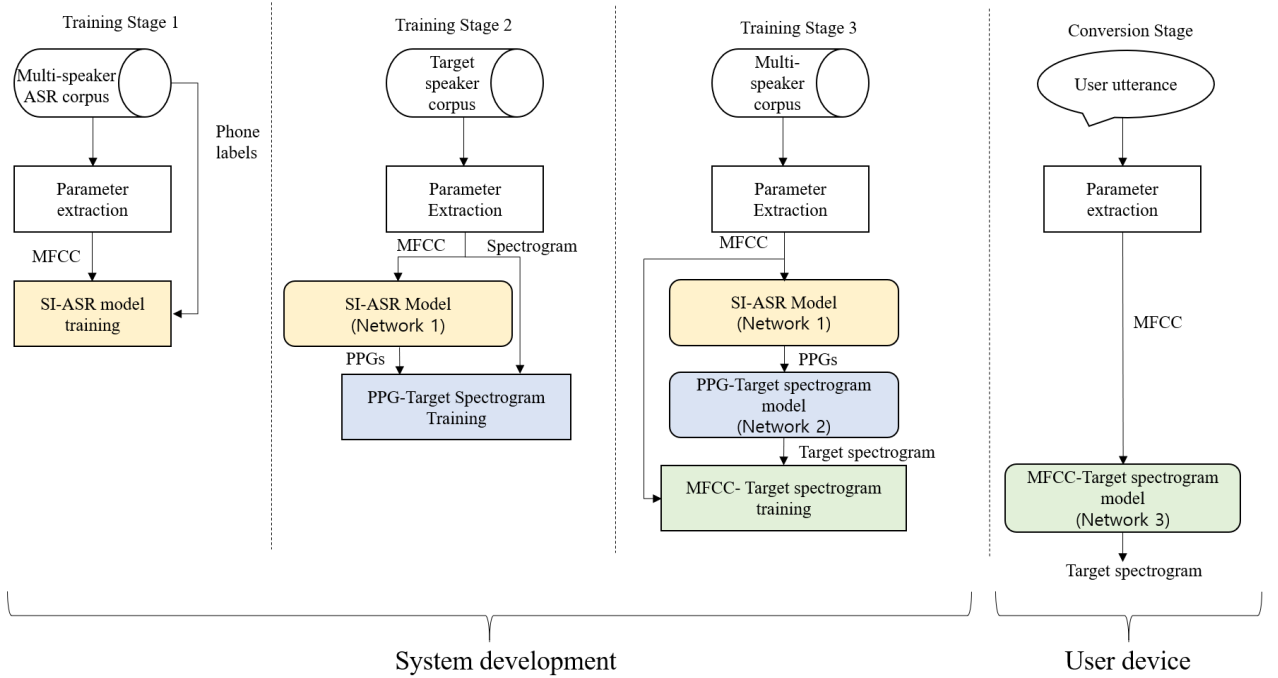

Figure 2: *Architecture of proposed non-parallel voice conversion.*

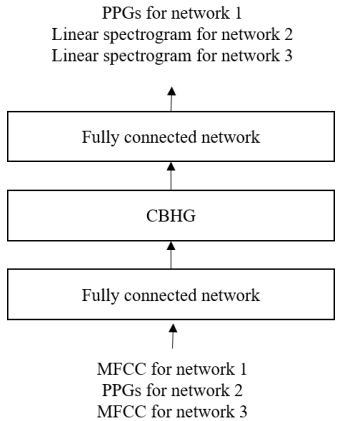

Figure 3: *Detailed implementation of each network.*

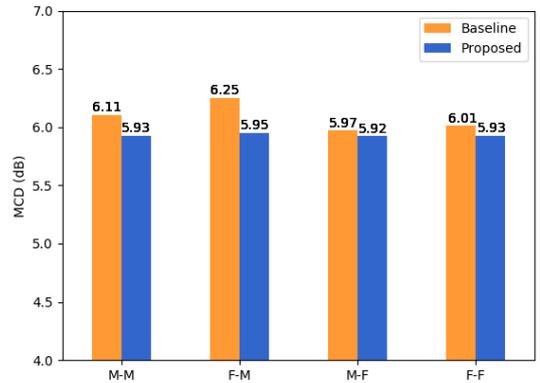

Figure 4: *MCD for baseline and proposed methods.*

for both baseline and the proposed approach. The equation for MCD is as in (3) [18],

$$MCD[dB] = \frac{10}{\ln 10}\sqrt{2\sum_{d=1}^{N}\left(c_d^{target} - c_d^{converted}\right)^2} \quad (3)$$

where N is the order of MFCC which is chosen as 40. $c_d^{target}$ represents target speaker MFCC at dimension d. $c_d^{converted}$ represents converted MFCC at dimension d. In this experiment, 10 utterances of 2 seconds length were averaged for each MCD.

In Figure 4, M and F stand for male and female speakers, respectively. Figure 4 shows that MCD values are higher when male is target. This tendency is also observed in [6]. We believe that this has to do with the fact that male speech has lower intelligibility in general. We assume that during training stage 2, the phone recognizer network produces higher estimation error for male and this contamination on the PPGs affects the performance of network 2 for the case where male is target.

Figure 4 shows that, for all gender pairs, the proposed method gives better MCD results. Especially for M-M and F-M conversion, the relative reductions are 2.95 % and 4.8 %, respectively, which can be significant improvement. The relative reduction is large for male target case. That is because the MCD is high for male target for the baseline while it has small variance for the proposed. The standard deviation is 0.12, 0.01 for the baseline and the proposed method respectively.

We believe the MCD reduction over all gender-pairs of the proposed method comes from the elimination of PPG estimation error at conversion stage. In other words, the linguistic bridging between source and target speech using PPG may result in some form of loss on phonetic information. The elimination of PPG estimation step during conversion reduces the dependency of conversion result on the intelligibility of individual source speaker.

Table 1. *Conversion time and number of network parameters for baseline and proposed method.*

| Models | Conversion time (seconds) | # of network parameters |
|---|---|---|
| Network 1 | 5.42 | 5,256,509 |
| Network 2 | 6.71 | 7,258,895 |
| Network 3 | 6.73 | 7,268,623 |
| Baseline (Network 1 + Network 2) | 12.13 | 12,515,404 |
| Proposed (Network 3) | 6.73 | 7,268,623 |
| Relative reduction (%) | 44.5 | 41.9 |

Table 1 describes the conversion time and the number of parameters required for baseline and the proposed method at the conversion stage. The conversion time was measured for 30 utterances, each with the length of 2 seconds. Because the time took for vocoding was the same for both cases, we excluded it from the conversion time. It can be seen from table 1 that the proposed method saves network parameters by 41.9 % and conversion time by 44.5 %.

## 5. Conclusions

In this paper, we propose a new non-parallel voice conversion approach that employs a single network for source-to-target feature mapping without the use of PPGs at conversion stage. Due to this straightforward architecture, our method runs faster and requires smaller amount of memory. It also improved the MCD by eliminating the information loss resulted from the PPG-based linguistic bridge. It is confirmed in experiments that the proposed method reduces the conversion time by 44.5 % and the amount of network parameters by 41.9 % with maximum 4.8 % MCD improvement compared to the baseline.

This reduction in conversion time by the proposed method can be further achieved when deployed in combination with network models of smaller size and fewer parameters.

As further work, it will be interesting to conduct performance comparison between the proposed approach and the parallel voice conversion method.

## 6. Acknowledgements

This material is based upon work supported by the Ministry of Trade, Industry & Energy (MOTIE, Korea) under Industrial Technology Innovation Program (No. 10080667, Development of conversational speech synthesis technology to express emotion and personality of robots through sound source diversification).